\documentclass[12pt, prd, showpacs]{revtex4}
\usepackage{amssymb}
\usepackage{amsmath}

\input{tcilatex}

\begin{document}

\title{Acceleration of particles as universal property of rotating black
holes}
\author{Oleg B. Zaslavskii}
\affiliation{Kharkov V.N. Karazin National University, 4 Svoboda Square, Kharkov, 61077,
Ukraine}
\email{zaslav@ukr.net}

\begin{abstract}
We argue that the possibility of having infinite energy in the centre of
mass frame of colliding particles is a generic property of rotating black
holes. We suggest a general model-independent derivation valid for dirty
black holes. The earlier observations for the Kerr or Kerr-Newman metrics
are confirmed and generalized.
\end{abstract}

\maketitle




\section{Introduction}

Quite recently, a series of works \cite{ban} - \cite{ted} appeared in which
interesting observations were made about energetics of particles near
rotating black holes. Namely, it was argued, that under certain conditions,
the energy in the center-of-mass frame can grow unbound, so a black hole
acts as a supercollider. This opens a window into a new physics including
the possibility of unknown channels of reaction between elementary
particles, with potential astrophysical applications such as elucidation of
the nature of active nuclei in the Galaxy \cite{agn}, etc. At the present,
these results were obtained for the Kerr metric and extended to the
Kerr-Newman one.

The aim of the present work is to show that this remarkable property of
being an accelerator to infinitely high energies is the direct consequence
of the general properties of the event horizon, provided one of colliding
particle approaches certain critical value of the angular momentum. We rely
not on the particular properties of the Kerr or Kerr-Newman \ metric but on
the generic axially symmetric rotating black holes. This is especially
important in the given context since physical significance of the effect
under discussion implies the presence of matter (say, accretion disc) around
the horizon, so the black hole, as usual in astrophysics, is "dirty". Thus,
our motivation is two-fold: to elucidate the essence of the effect from
general principles and to give derivation valid for black holes surrounded
by matter. The general approach which we push forward, enables to give
natural explanation to some important features of black holes as particles
accelerators, observed earlier in particular examples.

It was observed in \cite{jl} that the infinite acceleration can occur not
only for extremal black holes (as was stated in \cite{ban}) but also for
nonextremal ones, and the distinction between two cases was traced in detail
for the Kerr metric. This is important since in \cite{ban}, \cite{berti} 
\cite{ted} the effect under discussion was related to just extremal black
holes, meanwhile there are astrophysical limitations on the proximity of the
angular momentum of a black hole to the extremal value \cite{th}. In this
sense, the aforementioned result \ of \cite{jl} enables, in principle, to
evade this restriction and consider the effect not only for extremal black
holes. Therefore, it is desirable to trace whether this is retained for
astrophysically relevant "dirty" black holes. Now, different kinds of
limiting transitions and the role of the type of the horizon (nonextremal
versus extremal) follow directly from this general approach.

\section{Basic formulas and limiting transitions}

Consider the generic axially symmetric metric.\ it can be written as%
\begin{equation}
ds^{2}=-N^{2}dt^{2}+g_{\phi \phi }(d\phi -\omega dt)^{2}+dl^{2}+g_{zz}dz^{2}.
\label{z}
\end{equation}%
Here, the metric coefficients do not depend on $t$ and $\phi $. On the
horizon $N=0$. Alternatively, one can use coordinates $\theta $ and $r$,
similar to Boyer--Lindquist ones for the Kerr metric, instead of $l$ and $z$%
. In (\ref{z}) we assume that the metric coefficients are even functions of $%
z$, so the equatorial plane $\theta =\frac{\pi }{2}$ ($z=0$) is a symmetry
one.

In the space-time under discussion there are two conserved quantities $%
u_{0}\equiv -E$ and $u_{\phi }\equiv L$ where $u^{\mu }=\frac{dx^{\mu }}{%
d\tau }$ is the four-velocity of a test particle, $\tau $ is the proper time
and $x^{\mu }=(t,\phi ,l,z)$ are coordinates..The aforementioned conserved
quantities have the physical meaning of the energy per unit mass (or
frequency for a lightlike particle) and azimuthal component of the angular
momentum, respectively. It follows from the symmetry reasonings that there
exist geodesics in such a background which lie entirely in the plane $\theta
=\frac{\pi }{2}$. Then, the first integrals for such geodesics read (dot
denotes the derivative with respect to the proper time $\tau $):%
\begin{equation}
\dot{t}=u^{0}=\frac{E-\omega L}{N^{2}}.  \label{t}
\end{equation}%
We assume that $\dot{t}>0$, so that $E-\omega L>0$.%
\begin{equation}
\dot{\phi}=\frac{L}{g_{\phi \phi }}+\frac{(-\omega ^{2}L+E\omega )}{N^{2}},
\label{phi}
\end{equation}%
\begin{equation}
\dot{l}^{2}=\frac{(E-\omega L)^{2}}{N^{2}}-\delta -\frac{L^{2}}{g_{\phi \phi
}}.  \label{n}
\end{equation}

Here, $\delta =0$ for lightlike geodesics and $\delta =1$ for timelike ones.
For definiteness, we consider a pair of particles labeled by the subscript $%
i=1,2$ and having the equal rest masses $m_{1}=m_{2}=m$. We also assume that
both particle are approaching the horizon, so $\dot{l}<0$ for each of them.

The quantity which is relevant for us is the energy in the centre of mass
frame $E_{c.m.}=\sqrt{2}m\sqrt{1-u_{\mu (1)}u^{\mu (2)}}$\cite{ban} - \cite%
{ted}. After simple manipulations, one obtains from (\ref{t}) - (\ref{n})
that%
\begin{equation}
\frac{E_{c.m.}^{2}}{2m^{2}}=c+1-Y\text{, }c=\frac{X}{N^{2}}  \label{en}
\end{equation}%
where%
\begin{equation}
X=X_{1}X_{2}-Z_{1}Z_{2}\text{, }  \label{x}
\end{equation}%
$\,X_{i}\equiv E_{i}-\omega L_{i}$,%
\begin{equation}
Z_{i}=\sqrt{(E_{i}-\omega L_{i})^{2}-N^{2}b_{i}}\text{, }b_{i}=1+\frac{%
L_{i}^{2}}{g_{\phi \phi }}\text{,}  \label{z1}
\end{equation}%
\begin{equation}
Y=\frac{L_{1}L_{2}}{g_{\phi \phi }}\text{.}
\end{equation}%
Here, the crucial role is played by the quantity $c$ that determines whether
the energy can grow unbound. Now, we will discuss different limiting
transitions.

1) Let, for generic $L_{i}$, one approaches the horizon, so $N\rightarrow 0$%
. Expanding the radicals and retaining the first non-vanishing corrections
in the numerator, one obtains (subscript "H" refers to the horizon value):%
\begin{equation}
\left( \frac{E_{c.m.}^{2}}{2m^{2}}\right) _{H}=1+\frac{%
b_{1(H)}(L_{2(H)}-L_{2})}{2(L_{1H}-L_{1})}+\frac{b_{2(H)}(L_{(1)H}-L_{1})}{%
2(L_{2(H)}-L_{2})}-\frac{L_{1}L_{2}}{\left( g_{\phi \phi }\right) _{H}}\text{%
, }L_{i(H)}\equiv \frac{E_{i}}{\omega _{H}}.  \label{c}
\end{equation}%
By a very meaning of derivation, it is supposed in (\ref{c}) that $L_{1}\neq
L_{2(H)}$, $L_{2}\neq L_{2(H)}$.

Let us now specify the range of angular momenta in such a way that one of
them is close to the critical value: $L_{1}=L_{1(H)}(1-\varepsilon )$, $%
\varepsilon \ll 1$, $L_{2}\neq L_{2(H)}$. Then we have that%
\begin{equation}
\left( \frac{E_{c.m.}^{2}}{2m^{2}}\right) _{H}\approx \frac{%
b_{1(H)}(L_{2(H)}-L_{2})}{2L_{1(H})\varepsilon }\text{.}  \label{ecm}
\end{equation}%
This quantity can be made as large as one likes due to $\varepsilon
\rightarrow 0$. It follows from (\ref{en}) and (\ref{c}) that%
\begin{equation}
\lim_{L_{1}\rightarrow L_{1(H)}}\lim_{N\rightarrow 0}E_{cm}=\infty \text{.}
\label{1n}
\end{equation}

2) Let us take $L_{1}\rightarrow L_{1(H)}$ first and, then, consider the
limit $N\rightarrow 0$. The previous formula (\ref{c}) is valid both for
nonextremal and extremal horizons. In contrast to it, now the distinction
between two types of the horizon comes into play.

2a) First, consider the nonextremal case. We are interested in the immediate
vicinity of the horizon where the effect under discussion is expected to
show up. Near the horizon, we can infer the restriction that follows from
the condition of positivity of the expression inside the square root in (\ref%
{z1}). To this end, let us use the general form of the asymptotic expansion
for the metric coefficient $\omega $ that follows from the general
requirement of regularity of the geometry near the nonextremal horizon \cite%
{04}:%
\begin{equation}
\omega =\omega _{H}+BN^{2}+...  \label{ne}
\end{equation}%
Here, $\omega _{H}$ is constant and has the physical meaning of the angular
velocity of the horizon itself, the coefficient $B=B(\theta )$. For the case 
$\theta =\frac{\pi }{2}$ under consideration, $B$ is simply constant. Its
exact value is model-dependent.

Then, the condition of the positivity of (\ref{n}) cannot be satisfied since
the first term has the order $N^{2}$ whereas the others have the order $%
N^{0} $ and are negative. It means that the horizon is unreachable
(admissible region adjacent to the horizon shrinks to the point and there is
some turning point situated on a finite distance from the horizon).
Therefore, the present case should be rejected.

2b) Now, consider the extremal horizon. Then, instead of (\ref{ne}), one has
more general expansion%
\begin{equation}
\omega =\omega _{H}-B_{1}N+B_{2}N^{2}+...  \label{ex}
\end{equation}

The distinction between expansions for both horizons can be understood using
the Kerr metric as an example. The first corrections have the order $r-r_{H}$
where $r$ is the Boyer-Lindquist coordinate. However, for the nonextremal
case $N^{2}\sim r-r_{H}$ $\ $whereas for the extremal Kerr metric $N^{2}\sim
(r-r_{H}$ $)^{2}$, $B_{1}=M^{-1}$ where $M$ is the mass.

In a more general case, one can just appeal to the definition of the
nonextremal and extremal black holes using the proper length. Namely, in the
nonextremal case $N\approx \kappa l$ near the horizon where $\kappa $ is the
surface gravity and in the extremal one $N\approx N_{0}\exp (-Al)$, with $%
N_{0}$, $A=const>0$ and $l\rightarrow \infty .$ In principle, so-called
ultraextremal horizons with $N\sim l^{-s}$, $s>0$ are also possible which
can contain fractional powers of $r-r_{H}$ $\ $(where $r$ is the analogue of
the Boyer-Lindquist coordinate) but we do not discuss them here. (For the
spherically-symmetric configurations such horizons are classified in \cite%
{hor}.)

After the substitution of (\ref{ex}) to (\ref{x}), (\ref{z1}), we obtain
after simple manipulations that 
\begin{equation}
\frac{E_{c.m.}^{2}}{2m^{2}}\approx \frac{(E_{2}-\omega _{H}L_{2})}{N}[B_{1}%
\frac{E_{1}}{\omega _{H}}-\sqrt{(\frac{E_{1}^{2}}{\omega _{H}^{2}}%
B_{1}^{2}-b_{1})}]\text{.}  \label{extr}
\end{equation}%
Here, it is implied that the condition of the positivity is fulfilled for
the expression inside the radical (this cannot be worked out in more detail
in a model-independent way). Thus,%
\begin{equation}
\lim_{N\rightarrow 0}\lim_{L_{1}\rightarrow L_{1(H)}}E_{cm}=\infty \text{.}
\end{equation}

The extremal case has one more interesting feature. Namely, the proper time
needed to reach the horizon, tends to infinity. Indeed, it follows from (\ref%
{n}) and (\ref{ex}), that for the particle having $L=L_{(H)}$ and
approaching the horizon,%
\begin{equation}
\tau \sim \int \frac{dlN}{Z}\sim l\rightarrow \infty
\end{equation}%
since the proper distance from any point to the extremal horizon is
infinite. For the nonextremal horizon the proper distance is finite as well
as the proper time. Also, one can easily find from (\ref{phi}), (\ref{n})
that the number of revolutions%
\begin{equation}
\Delta \phi \approx \frac{EB_{1}}{\sqrt{(\frac{E^{2}}{\omega _{H}^{2}}%
B_{1}^{2}-b_{1})}}\int \frac{dl}{N}  \label{rev}
\end{equation}

Using again the asymptotic form $N\approx N_{0}\exp (-Al)$ in the extremal
case, we see that $\Delta \phi \rightarrow \infty $.

3) In two previous situations, the result was formally determined by\ a play
between two small quantities $\varepsilon $ and $N$ and the order of taking
the limits $\varepsilon \rightarrow 0$ and $N\rightarrow 0$. Meanwhile, it
is of interest to trace, what happens when both quantities are small but
nonzero, and what are limitations on the possibility of collision with
infinitely growing energies. For the Kerr metric, a particle with the
critical value of the angular momentum $L_{(H)}$ cannot come from infinity
since a potential barrier prevents this, so that the energy cannot grow
unbound as a result of single scattering \cite{ban}, \cite{ted}. Meanwhile,
as was demonstrated for the Kerr metric \cite{jl}, this becomes possible if
multiple scattering occurs, so one of colliding particles does not come from
infinity but receives the near-critical angular momentum as a result of
collision near the horizon. As far as the generic spacetime is concerned, in
principle it can happen that a particle with the critical angular momentum
coming from infinity is able to reach the horizon. However, such a
possibility is model-dependent and requires special conditions on the
behavior of the metric. Meanwhile, we are interested in features that have
general model-independent character. Therefore, we will not discuss such
particular cases and assume that a particle has a near-critical angular
momentum in the near-horizon region just due to multiple collisions. Let us
see the necessary condition for this.

From the condition $Z^{2}\geq 0$ where $Z$ is defined in (\ref{z1}) we
obtain for the nonextremal horizon using (\ref{ne}) that the process under
discussion can indeed occur but only in the narrow strip near the horizon
where%
\begin{equation}
0\leq N\leq \frac{E\varepsilon }{\sqrt{b_{H}}}\text{.}  \label{N}
\end{equation}

Thus, the energy (\ref{ecm}) can be indeed as large as one likes but this
happens provided a particle acquires the near-critical angular momentum in
the region bounded by eq. (\ref{N}). If $\varepsilon =0$ exactly, the
permitted strip shrinks to the point and we returned to case 2a when the
effect is impossible. For the extremal horizon, there is no limitation
similar to (\ref{N}) since one can put $\varepsilon =0$ in accordance with
case 2b), see also eq. (\ref{ex}).

4) For completeness, let us consider the case when simultaneously $%
L_{1}=L_{1(H)},L_{2}=L_{2(H)}$. In the nonextremal case it follows from (\ref%
{ne})\ that the radical cannot remain positive near the horizon and the
horizon is unreachable, so this case is irrelevant for our analysis. For the
extremal one, in the horizon limit we obtain from (\ref{ex}) that the terms
of the order $N$ in the numerator cancel, so that the first non-vanishing
term has the same order $N^{2}$ as the denominator. As a result, the
quantity $E_{c.m.}$ is finite. However, the proper time needed to reach the
horizon is still infinite.

If we compare the meaning of limits 1 and 2, we see that in the nonextremal
case the energy in the centre of mass frame is finite but can be made as
large as one likes if the angular momentum of one of two colliding particles
is chosen arbitrarily close to the critical value. If this value is chosen
exactly equal to the critical value from the very beginning, the energy can
be made as large as one wishes when one approaches the horizon (it becomes
possible in the extremal case only). This is direct generalization of
observations made in \cite{ban}, \cite{jl} for the Kerr metric.

\section{Comparison of general results with case of Kerr metric}

It is instructive to compare some general results obtained in our paper to
those obtained earlier for the Kerr metric. Then, for the equatorial plane $%
\theta =\frac{\pi }{2}$,%
\begin{equation}
g_{00}=-(1-\frac{2M}{r})\text{, }g_{0\phi }=-\frac{2Ma}{r}\text{, }g_{\phi
\phi }=r^{2}+a^{2}+\frac{2Ma^{2}}{r}\text{, }\omega =-\frac{g_{0\phi }}{%
g_{\phi \phi }}\text{.}
\end{equation}%
\begin{equation}
N^{2}=\frac{(r-r_{H})(r-r_{C})}{r^{2}+a^{2}+\frac{2M}{r}a^{2}}\text{,}
\end{equation}%
where $r$ is the Boyer-Lindquist coordinate, $r_{H}=M+\sqrt{M^{2}-a^{2}\text{
}}$, $r_{C}=M-\sqrt{M^{2}-a^{2}\text{ }}$, the horizon value of the
coefficient $\omega $ is equal to $\omega _{H}=\frac{a}{2Mr_{H}}$. If we
define $L=lEM$ and take $E=1$ (that corresponds to a particle falling from
infinity from the state of the rest), the critical value of the angular
momentum, $l_{(H)}=\frac{2r_{H}}{a}$. Then, using (\ref{ne}) - (\ref{c}) one
can calculate the energy in the centre of mass for a collision on the
horizon:%
\begin{equation}
\left( \frac{E_{c.m.}}{2m}\right) _{H}=\sqrt{1+\frac{M(l_{1}-l_{2})^{2}}{%
2r_{C}(l_{1}-l_{H})(l_{2}-l_{H})}}  \label{ekerr}
\end{equation}%
that coincides exactly with eq. (10) of \cite{jl} from which further
analysis of collisions in the Kerr metric can be carried out.

Near the horizon $b\approx \frac{2Mr_{H}}{a^{2}}$ and eq. (\ref{N}) turns
into%
\begin{equation}
0\leq r-r_{H}\leq \frac{a^{2}\varepsilon ^{2}}{r_{H}\sqrt{1-\frac{a^{2}}{%
M^{2}}}}  \label{h}
\end{equation}%
that is completely equivalent to eq. (18) of Ref. \cite{jl} where instead of 
$\varepsilon $ the quantity $\delta =\varepsilon \frac{2r_{H}}{a}$ was used.
A particle with $E=1$ cannot penetrate from infinity to the horizon but,
nonetheless, there is a narrow region between a horizon a potential barrier
where such motion can occur that can generate acceleration to arbitrarily
large energies (see \cite{jl} for details). Eq. (\ref{ekerr}) is valid for
both nonextremal and extremal cases, eq. (\ref{h}) applies to the
nonextremal metric. In the extremal case eq. (\ref{extr}) turns into a very
simple formula%
\begin{equation}
\frac{E_{c.m.}^{2}}{2m^{2}}\approx \frac{(2-l_{2})}{2N}(2-\sqrt{2})\text{.}
\label{sim}
\end{equation}%
It can also be obtained from eq. (17) of \cite{gp3} by putting there $%
\varepsilon =1$, $l_{H}=2$.

In the case $E=1$, substituting $B_{1}=M^{-1}$, $\omega _{H}=(2M)^{-1}$, $%
r-M $ $\approx M\exp (-\frac{l}{M})$, $N\approx \frac{r-M}{2M}$ into (\ref%
{rev}), we obtain that $\Delta \phi $ $\approx \frac{M\sqrt{2}}{r-M}%
\rightarrow \infty $ \ in agreement with \cite{jl}.

Thus, our general results from the previous section correctly reproduce the
basic formulas for the Kerr metric.

\section{Discussion and conclusions}

Thus, we suggested very simple and direct derivation of the effect of
growing energy from first principles and without using the explicit form of
the black hole metric. This became possible due to the fact that the
relevant region is the vicinity of the horizon only where universality of
the black hole physics reveals itself. In particular, we generalized recent
observations made for the Kerr metric in \cite{jl} and showed that,
generically, for the nonextremal rotating black hole the horizon value of
the energy in the centre of mass is finite but can be made as large as one
likes if the angular momentum of one colliding particle approaches the
critical value. In the \ extremal case, the energy for the critical value of
the momentum grows unbound as a horizon is approached but the proper time
also so does. In this respect, the mechanism preventing infinite energies
has an universal character.

It was stated that there are astrophysical limitations on the significance
of the effect in question due to gravitational radiation, backreaction, etc. 
\cite{berti}, \cite{ted}. We did not consider here the role of such
mechanisms having restricted ourselves by the picture of geodesic motion. A
separate important task beyond the scope of the present paper that needs
further attention is to evaluate the relative role of such effects that also
includes studying some concrete models. There is one more obstacle to get
infinitely large energies in the nonextremal case since, generically, one
particle should acquire the critical angular momentum in the very narrow
strip near the horizon due to multiple scattering only. Therefore, further
investigation of these issues is needed. Nonetheless, bearing in mind that
the main results described above have the universal character, potential
acceleration to large (formally, infinite) energies should be taken
seriously both as manifestation of general properties of black holes and the
effect relevant in astrophysics of high energies.

I thank Yuri Pavlov for the stimulating interest to this work and fruitful
discussions.

\end{document}